# Local distortion in $LaCoO_3$ and $PrCoO_3$: EXAFS, XRD and XANES studies

S. K. Pandey[1], S. Khalid[2], N. P. Lalla[1], and A. V. Pimpale[1]

[1]UGC-DAE Consortium for Scientific Research, University Campus, Khandwa Road, Indore 452 017, India

[2]National Synchrotron Light Source, Brookhaven National Laboratory, Upton, New York – 11973

E-mail: sk_iuc@rediffmail.com and avp@csr.ernet.in

**Abstract**

Room temperature Co K-edge extended x-ray absorption fine structure (EXAFS), x-ray absorption near edge structure (XANES) including pre-edge and x-ray diffraction (XRD) studies are carried out on $LaCoO_3$ and $PrCoO_3$. The Co-O, Co-La/Pr and Co-Co bond lengths are obtained from EXAFS analysis and compared with those obtained from XRD. The EXAFS analysis of data indicates that $CoO_6$ octahedron is distorted in both $LaCoO_3$ and $PrCoO_3$. There are two Co-O bonds with bond length 1.863 (1.886) Å and four Co-O bonds with bond length 1.928 (1.942) Å for $LaCoO_3$ ($PrCoO_3$). Such distortion is expected in orthorhombic $PrCoO_3$ but not in rhombohedral $LaCoO_3$. This distortion in $CoO_6$ octahedron is attributed to Jahn-Teller active $Co^{3+}$ ion in intermediate spin state in these compounds. Higher shell studies reveal that Debye-Waller (DW) factors of Co-Pr and Co-Co bonds in $PrCoO_3$ are more in comparison to Co-La and Co-Co bonds in $LaCoO_3$ indicating that these bonds are structurally more disordered in $PrCoO_3$. The comparison of Co-Co bond lengths and corresponding DW factors indicates that the structural disorder plays an important role in deciding the insulating properties of these compounds. XANES studies have shown changes in the intensities and positions of different near edge features. The comparison of experimental spectra with the calculated ones- using Co 4p density of states obtained from local density approximation calculations and matrix elements calculated using atomic like core state as initial and confluent hypergeometric function as final states- indicates that for orthorhombic structure intensities of different features are lower as compared to the cubic structure. The pre-edge peaks attributed to Co 1s $\to e_g^\uparrow$ and $e_g^\downarrow$ transitions show effect of hybridization of the $e_g$ orbitals with O 2p orbitals and their relative intensities in $PrCoO_3$ and $LaCoO_3$ are explained by using average Co-O bond length obtained from EXAFS.

**Introduction**

Cobaltates with general formula $ACoO_3$ (A-rare earth element) is an interesting class of compounds in the perovskite family. They show temperature induced insulator to metal and non-magnetic to paramagnetic transitions [1-3]. The transport and magnetic properties of these compounds are dependent upon the spin-state of the $Co^{3+}$ ion, which is decided by the competition between the Hund's coupling energy and the crystal-field splitting energy [4]. These energies are comparable within a few meV and so thermal excitations can readily affect the occupancies of different $e_g$ orbitals thereby affecting the spin state of the $Co^{3+}$ ion. As temperature increases these compounds show two main magnetic transitions- one at low temperatures around 100 K for $LaCoO_3$ and 200 K for $PrCoO_3$ and the other at high temperatures around 500 K for $LaCoO_3$ and 550 K for $PrCoO_3$. Earlier the high temperature transition was generally attributed to a high spin (HS) $t_{2g}^4 e_g^2$ state and the low temperature transition to an admixture of HS and low spin (LS) $t_{2g}^6 e_g^0$ states [5]. However, later on Potze et al. introduced the concept of intermediate spin (IS) $t_{2g}^5 e_g^1$ state in these compounds [6] and attributed the low temperature transition to LS to IS transition. Currently, it is generally believed that the spin-state of $Co^{3+}$ ion changes from LS→IS→HS with increase in temperature.

The spin-state of $Co^{3+}$ ion is closely related with the structural parameters of these compounds. It is well known that the ionic radius of an ion depends on its spin-state. The lattice parameters of a perovskite depend on its tolerance factor, which is directly related with the ionic radii of various ions. Therefore, any change in spin-state will directly affect the lattice parameters and thereby the unit cell volume, metal-oxygen bond lengths, etc. [7]. At room temperature for both the compounds under study the $Co^{3+}$ ion is in IS state [8,9]. In the IS state $e_g$ orbital possesses one electron making $Co^{3+}$ Jahn-Teller (JT) active and it should influence the Co-O distances in the $CoO_6$ octahedron and distort it. Such distortion has been seen in manganites where JT effect is important [10-12]. One expects similar structural distortion in the case of cobaltates with $Co^{3+}$ in IS state. The x-ray and neutron powder diffraction studies of $LaCoO_3$ have shown the system to be rhombohedral at room temperature and as such the crystal symmetry demands the Co-O bond lengths to be identical for all the six bonds in the $CoO_6$ octahedron. To resolve the Co-O bond lengths through diffraction one needs to see the splitting of Bragg peaks which is limited by the experimental resolution. The pulsed neutron pair distribution function (PDF) studies have shown the existence of two Co-O bond lengths at 100 K [13]. For $PrCoO_3$, the x-ray structure is orthorhombic and one expects that the Co-O bond lengths need not be the same. There is no neutron data on this compound to the best of our knowledge.

Extended x-ray absorption fine structure (EXAFS) observed about 50 eV above the main-edge of the x-ray absorption (XA) spectrum contains information about the local structure around the absorbing/central atom. By analyzing the EXAFS data one can get the information about the bond lengths of the neighbouring atoms with respect to the central atom. The accuracy of the bond lengths determined from EXAFS data depends upon the photon energy range over which XA data is taken. Higher this range- larger would be the photoelectron wave vector k-range and better would be the ability to distinguish two very similar bond lengths. Thus high k-range EXAFS data offer better opportunity to resolve bond lengths which usual diffraction experiments are unable to do. Thus it is interesting to carry out EXAFS studies of cobalates and compare the same with the x-ray diffaction studies.

The x-ray near-edge structures (XANES) are sensitive to the electronic states of the absorbing atoms and the local structure through distribution of neighbouing atoms [14-17]. Recently we have studied the Co K- and La L-edges of $LaCoO_3$ and attributed different near edge features to the overlap of atomic orbitals of absorbing atoms to that of neighbouring atoms. The overlap of these orbitals is sensitive to the bonds lengths. Therefore one expects changes in XANES with change in the bond lengths. The changes in different near edge features in the doped compounds are generally attributed to change in local and electronic structures. The separation of these two contributions is rather difficult as substitution of an atom by another atom changes the structure as well as the electronic states. However, the compounds under study seem to be good candidates to see the effect of structure on the XANES as the electronic state of Co ion is same at room temperature. Change in the rare-earth site only changes the structural parameters due to change in the ionic radius without affecting much the electronic state of Co ions. Thus, simultaneous XANES and EXAFS studies of these compounds will throw some light on the influence of local structure on XANES.

We report here the room temperature Co K-edge EXAFS, XANES and XRD studies of $LaCoO_3$ and $PrCoO_3$. The EXAFS results have shown distortion in $CoO_6$ octahedron for both $LaCoO_3$ with rhombohedral and $PrCoO_3$ with orthorhombic structures. This distortion is attributed to Jahn-Teller active $Co^{3+}$ ion in intermediate spin state in these compounds. The Debye-Waller factors of higher shells reveal that the $PrCoO_3$ has more structural disorder than $LaCoO_3$. XANES studies of these compounds show that the orthorhombic distortion decreases the intensity of the main-edge and post edge peaks. However, the intensity of pre-edge peaks for orthorhombic $PrCoO_3$ is seen to increase due to decreased overlap of Co $e_g$ orbitals with O 2p orbitals as revealed by increase in average Co-O bond length obtained from EXAFS.

**Experiment and data analysis**

The polycrystalline samples of $LaCoO_3$ and $PrCoO_3$ were prepared by combustion method and characterized by x-ray diffraction (XRD) and resistivity techniques. The details of sample preparation and characterization are given in our earlier publications [17-19]. Room temperature Co K-edge XAS experiments were done at beamline X-18 B at the National Synchrotron Light Source, Brookhaven National Laboratory. The storage ring was operated at 2.8 GeV, 300 mA. The beamline used a Si (111) channel cut monochromator. The horizontal acceptance angle of the beam at the monochromator was 1 mrad. The vertical slit size used in this experiment was 1 mm, corresponding to an energy resolution of about 0.8 eV at the Co K-edge. The average photon flux for this bandwidth was $10^{10}$ photons/sec. The monochromator was detuned by 35 % to reduce the higher harmonics. The incident ($I_0$) and the transmitted beam ($I_t$) were measured by sealed ion chambers, with a combination of gases for appropriate absorption. Standard Co foil was placed between the detectors $I_t$ and $I_{ref}$ for energy reference and to check the stability of the beamline and optical system. The samples sieved through a 400 mesh were spread uniformly on a cellophane tape and different layers of this tape were used to minimize the pinhole and brick effects. The pinhole corresponds to absence of the sample material and the brick corresponds to increased thickness of the absorber material in the path of the incident radiation. The value of $\mu x = \ln(I_0/I_t)$ would thus be different in the pinhole and the brick areas of the sample, for the former it would be almost zero and for the latter it would be very large. Such inhomogeneity distorts the recorded spectra [20]. We have used 2, 3, and 4 layers of the sample adhered cellophane tape and recorded the spectra. The Fourier analysis of the data for all the three thicknesses showed exactly the same results indicating the uniformity of the sample thickness. The data given in this work correspond to 4 layers of the tape and the values of μx are in the range 0.87 to 1.4 for $LaCoO_3$ and 1.7 to 2.4 for $PrCoO_3$ around the main-edge. The x-ray powder diffraction patterns were recorded with monochromatised Cu-$K_\alpha$ radiation in the 2θ range of 10° - 90° using Rigaku powder x-ray diffractometer. The rotating anode x-ray generator was operated at 40 kV and 100 mA. The monochromator used was graphite (002) and the widths of the divergent slit, scattering slit and the receiving slit were 0.5°, 0.5°, 0.15 mm, respectively. The data were collected with a step size of 0.02° with a scanning rate of 2° per minute. To obtain the lattice parameters and atomic positions XRD patterns were fitted using Rietveld profile refinement technique [21]. The data were fitted by considering rhombohedral (R-3c) and orthorhombic (Pnma) structures for $LaCoO_3$ and $PrCoO_3$, respectively. The final refined lattice parameters match well with those reported in the literature. These refined lattice parameters and atomic positions were used as input parameters for "ATOM" program of FEFF group to estimate the Co-O, Co-La/Pr and Co-Co bond lengths. Finally, the errors in the bond lengths were estimated by changing these input parameters by the error amount in their values as yielded by the Reitveld program.

EXAFS fitting was carried out by using UWXAFS 3.0 software [22]. The threshold energy, $E_0$, for both the spectra was taken as the first inflection point in the absorption edge region. After the background subtraction, the absorption coefficient μ(E) was converted to μ(k), where k = $(2m(E-E_0)/\hbar^2)^{1/2}$ is the magnitude of wave vector of the ejected photoelectron. The EXAFS oscillation χ(k) is defined as, $(\mu-\mu_0)/\mu_0$, where $\mu_0$ is the embedded atom absorption coefficient. The Fourier transform (FT) to the r-space was taken in the k range 3-14 $Å^{-1}$ by Fourier transforming $k^2\chi(k)$ with Hanning window. To find the local distortion in $CoO_6$ octahedron first shell fitting was done in the Fourier filtered k space in the range 0.65 to 1.90 Å for $LaCoO_3$ and 0.6 to 1.9 Å for $PrCoO_3$. We have used different model structures to get the information about the distortion in $CoO_6$ octahedron. The average structure inferred from x-ray diffraction for $LaCoO_3$ is rhombohedral implying that all the six Co-O bonds of CoO$_6$ octahedron of equal length. First, we tried *6 model* consisting of only one O-shell with six equal Co-O bond lengths. As mentioned above Louca et al. [10] have reported the distorted CoO$_6$ octahedra at local level with more than 4 short and less than 2 long bond lengths at 100 K in IS spin-state. Therefore, we also modeled local structure using *4+2 model* consisting of two-shells with 4 equal short bond lengths and 2 equal long bond lengths. According to XRD results for PrCoO$_3$ the structure is orthorhombic and there are three Co-O bond lengths each corresponding to a pair of bonds: 1.901, 1.931, and 1.947 Å. As the last four bonds are nearly equal, hence in addition, we used *2+4 model*, which consists of two-shell fitting with two equal short bond lengths and four equal long bond lengths. The higher shell fitting was also carried out to get the information about Co-La/Pr and Co-Co bond lengths. During this fitting the Co-O bond lengths were kept fixed to their values obtained from two-shell fitting over CoO$_6$ octahedron. To get reliable information about these bond lengths we have not considered the distribution of bonds in each shell and only single scattering was considered as such considerations would enhance the number of variable parameters and corresponding paths. Thus we have concentrated in getting the average bond lengths of each higher shell. For this purpose we have carried out the higher shell fitting by considering two-shell for 8 Co-La/Pr bonds and one-shell for 6 Co-Co bonds in the Fourier filtered k space in the range 0.65 to 3.85 Å for LaCoO$_3$ and 0.6 to 3.9 Å for PrCoO$_3$. The overall many body reduction factor $S_0^2$ arises from overlap contribution of the passive electrons. It is typically around 0.7 to 0.9 and it is not very sensitive to chemical environment. For La$_{1-x}$Sr$_x$MnO$_3$, the value of $S_0^2 = 0.82$ was used in our earlier work [11] as it gave best fit. We have adopted the same value for both LaCoO$_3$ and PrCoO$_3$ as replacing Mn by Co is not expected to change this value significantly. Our results also show that the chosen value of $S_0^2 = 0.82$ gave best fit. The back scattering amplitude and phase shifts were calculated using FEFF6.01 [23]. During the fitting the number of atoms in the different coordination shells (N) was kept fixed as per the choice of model structure and only bond lengths and corresponding Debye-Waller factors ($\sigma^2$) were varied to reduce the number of correlated parameters. Further, several different fits were performed in each case to verify the robustness of the parameters. Both-simultaneous variation of different fitting parameters as well as their independent variation was tried. A fit is considered to be good if the goodness of fit parameter given by R-factor is less than 0.02 [22]. Since for each compound fitting was tried using different structure models, the model that resulted in the least R-factor was considered to represent best the local structure for that particular compound. The bond lengths and the errors in their values are yielded directly by the fitting program when bond lengths are varied for the fitting purposes.

**Computational details**

Spin polarized local density approximation (LDA) band structure calculations of $LaCoO_3$ were carried out using LMTART 6.61 [24]. The rhombohedral $LaCoO_3$ is modeled for simplicity by a cubic perovskite structure with lattice parameters corresponding to double the Co-O bond length obtained from XRD. To see the effect of lattice distortion on XANES we have also considered orthorhombically distorted perovskite structure with lattice parameters taken as double of Co-O bond length obtained from XRD for $PrCoO_3$. For calculating charge density, full-potential LMTO method working in plane wave representation was used. The muffin-tin radii used in the calculations are 3.509, 2.001, and 1.637 au for La, Co, and O sites, respectively. The charge density and effective potential were expanded in spherical harmonics up to $l = 6$ inside the sphere and in a Fourier series in the interstitial region. The initial basis set included 6s, 6p, 5d, and 4f valence and 5s, 5p semicore orbitals of La; 4s, 4p, and 3d valence and 3p semicore orbitals of Co and 2s and 2p valence orbitals of O. The exchange correlation functional of the density functional theory was taken after Vosko, Wilk, and Nussair [25]. (6, 6, 6) divisions of the Brillouin zone along three directions for the tetrahedron integration were used to calculate the DOS. Self-consistency was achieved by demanding the convergence of the total energy to be smaller than $10^{-5}$ Ryd/cell.

Under dipole approximation the K-edge absorption process occurs due to transition of electron from s-symmetric state to p-symmetric unoccupied states. Due to participation of localized states in transition the contribution of states of atoms other than the Co to the Co K-edge absorption is negligible. Thus Co K-edge absorption can be identified to the electronic transition from Co 1s to 4p states. To calculate the absorption spectra we need matrix elements and DOS of Co 4p states. We have calculated the matrix element $\langle \psi_F | H_{int} | \psi_I \rangle$ by using $|\psi_I\rangle$ as the 1s hydrogenic wave function and $|\psi_F\rangle$ as confluent hypergeometric function [26]. Matrix element thus obtained has been used to obtain the transition probability per second ($w$). To calculate the absorption spectrum, calculated $w$ is convoluted with a Lorentzian having FWHM of 2.1 eV to account for the lifetime of Co 1s core hole generated during the absorption process [27] and a Gaussian having FWHM of 2 eV to account for the instrumental and other brodening.

**Results and discussion**

### EXAFS

Figure 1 shows the $k^2$ weighted Co K-edge EXAFS spectra $\chi(k)$ for LaCoO$_3$ and PrCoO$_3$. The k-range used for analysis is from 3-14 Å$^{-1}$. Both the spectra are highly structured as expected from the powder crystal nature of the samples. As one goes from LaCoO$_3$ to PrCoO$_3$ there is appreciable change in spectrum indicating changes in the local structure around Co sites. The FT of both the $k^2\chi(k)$ spectra are shown in figure 2. These spectra are uncorrected for the central and back-scattered phase shifts. We have fitted the spectra in the r-range 0.6 to 3.9 Å; for r < 0.6 Å the data reveals structure due to limited k-range and for r > 3.9 Å one has to go higher coordination shells. In this range for LaCoO$_3$ we observe three main peaks marked by 1, 2 and 3 ascribed to the first, second and third coordination shells of the central Co atom comprising of the 6 O-atoms of the CoO$_6$ octahedron, 8 La-atoms and 6 Co-atoms, respectively as per expectations from perovskite structure. For PrCoO$_3$ we observe four main peaks denoted by 1, 2′, 2, and 3. Peaks 1 and 3 correspond to Co-O and Co-Co shells. Peak 2 corresponding to second coordination shell of central Co atom surrounded by 8 La-atoms in LaCoO$_3$ is well split into two peaks in the PrCoO$_3$ denoted by 2′ and 2. The intensities of the peaks 2′, 2 and 3 are significantly lower for PrCoO$_3$ indicating the increased local disorder for second and third shells in comparison to that for LaCoO$_3$. The XRD results revealed rhombohedral structure for LaCoO$_3$ and orthorhombic structure for PrCoO$_3$. In the rhombohedral structure out of 8 Co-La bonds two are at 3.273 Å and remaining six are at 3.325 Å, whereas for orthorhombic structure the 8 bonds are at 3.140, 3.250, 3.333, and 3.411 Å with two each at the same distance. Similarly, all the 6 Co-Co bonds are exactly at equal distance (3.825 Å) in the rhombohedral structure, whereas in the orthorhombic they split into two groups with 2 bonds at 3.787 Å and remaining 4 at 3.789 Å. Thus one expects more disorder in the $PrCoO_3$ due to its orthorhombic structure. The two peaks α and β observed between peaks 1 and 2 (2′ for $PrCoO_3$) are attributed to spectral leakage and double scattering peaks corresponding to Co-O-O-Co.

The single- and double-shell fits over $CoO_6$ octahedron for $LaCoO_3$ and $PrCoO_3$ are shown in figures 3 and 4, respectively. It is clear from the figures that the *6 model* which considers all the 6 Co-O bonds at the same distance does not give good fit at higher k values. On considering *2+4 model*- where 2 short bonds are at same distance and remaining 4 bonds are long and at the same distance- the fit improves. We have also tried *4+2 model*, which considers 4 short bonds at the equal distance and remaining 2 are long and at the same distance, however, this model does not improve the fit. Thus, *2+4 model* is a better representative of the experimental data. Parameters obtained from the fitting are given in table 1. It is clear from the table that the difference between the long and short bonds is 0.065 Å for $LaCoO_3$ where as for $PrCoO_3$ it is 0.056 Å. This difference is well beyond the accuracy 0.008 Å of the bond length. Thus, $LaCoO_3$ is slightly more distorted at the local level. The distortion in the $CoO_6$ octahedron at local level is interesting for $LaCoO_3$ as its structure is rhombohedral. In rhombohedral structure one would expect all the six Co-O bonds of equal length. Such distribution in bond length of $CoO_6$ octahedron at local level may be a direct evidence of JT distortion expected in these compounds with $Co^{3+}$ ions in IS state. In this state one electron goes to $e_g$ level making $Co^{3+}$ ion JT active. Louca et al. [10] and Maris et al. [28] have also seen distortion in $CoO_6$ octahedron in $LaCoO_3$. Louca et al. have carried out temperature dependent PDF studies using pulsed neutron on a powder sample. They have seen two peaks at 1.915 and 2.16 Å corresponding to two bond lengths at 100 K. The area under these peaks gives information about the number of equal bonds and sum of these areas is normalized to six; the number of short bonds is greater than four and the number of long bonds correspondingly is lower than two. They have seen the decrease in the area of peak at 2.16 Å with increase in temperature and this peak merges almost to the background at 300 K. The experiment of Maris et al. dealt with high-resolution single crystal XRD studies with synchrotron radiation. They noted monoclinic distortion with three different Co-O bond lengths. Contrary to Louca et al. the distortion increased to a higher value as temperature increased to room temperature. Our room temperature EXAFS studies also reveal that there is a distortion in $CoO_6$ octahedron and the nature of distortion is different from that obtained from Louca et al. [10] at 100 K. In our case there are 2 short bonds at 1.863 Å and 4 long bonds at 1.928 Å. It is interesting to note that these short and long bond lengths are closer to the short and middle bond lengths of Maris et al. at 295 K. They have attributed this distortion to cooperative JT effect.

The higher shell fittings have been performed on $LaCoO_3$ and $PrCoO_3$ data to find the average Co-La/Pr and Co-Co bond lengths; the Co-O bond lengths having kept fixed to the values obtained from first shell fitting. These fits are plotted in figures 5 and 6 for $LaCoO_3$ and $PrCoO_3$, respectively. It is clear from the figure that the fitted data deviate from the experimental ones at higher k indicating that there is a distribution in these bond lengths as well at the local level. The deviation is more for $PrCoO_3$ suggesting that the distribution in bond lengths is more in this compound. Parameters obtained from fitting are given in tables 1 and 2. Bond lengths obtained from XRD are also given in these tables. It is seen that for $LaCoO_3$ there are six Co-O bonds at average 1.906 Å, two Co-La bonds at 3.251 Å and remaining six Co-La bonds at 3.334 Å and six Co-Co bonds at 3.894 Å. For $PrCoO_3$ there are six Co-O bonds at average 1.923 Å, two Co-Pr bonds at 3.128 Å and remaining six Co-Pr bonds at 3.308 Å and six Co-Co bonds at 3.818 Å. This indicates that the average Co-O bond length in $PrCoO_3$ is more than that in $LaCoO_3$ whereas other bond lengths are small in comparison to $LaCoO_3$. Here it is important to note that the difference between two Co-Pr bonds is 0.18 Å which is more than the EXAFS resolution $\Delta R = \pi/[2 (k_{max} - k_{min})] \approx 0.143$ Å and we were able to see experimentally well resolved two peaks 2′ and 2 corresponding the 2 and 6 Co-Pr bonds, respectively. The Debye-Waller (DW) factor for Co-Pr bond is higher than the corresponding factor for Co-La bond indicating higher disorder in $PrCoO_3$ as compared to $LaCoO_3$. As thermal contribution to the disorder would be almost the same in both cases, it may be inferred that the structural disorder is more for Co-Pr bonds as compared to Co-La bonds. Similarly the DW factor corresponding to Co-Co bonds is more in $PrCoO_3$ suggesting that this shell is also more disordered at local level in $PrCoO_3$ in comparison to LaCoO$_3$. This result is quite interesting in the light of insulating behaviour of these compounds as PrCoO$_3$ is more insulating than LaCoO$_3$ [29]. The insulating property is governed by electron hopping probability from one Co-site to the next Co-site and higher the bond length lower the overlap of electronic wavefunctions and smaller the hopping probability leading to higher resistivity. The Co-Co bond length in PrCoO$_3$ is less, therefore one expects PrCoO$_3$ to be less insulating than LaCoO$_3$ contrary to the experimental situation [29]. Thus it seems that besides bond length the higher structural disorder in PrCoO$_3$ plays an important role in the transport behaviour.

The comparison of the bond lengths obtained from EXAFS and XRD is quite interesting. It is seen from table 1 that XRD Co-O bond length is higher in LaCoO$_3$ as compared to that in PrCoO$_3$, whereas the EXAFS Co-O bond length is higher in PrCoO$_3$. The XRD bond lengths obtained from powder data are constrained by the assumed model of the crystal structure- rhombohedral for LaCoO$_3$ and orthorhombic for PrCoO$_3$- whereas EXAFS analysis permitted existence of distortion in CoO$_6$ octahedron. The validity of EXAFS results is reinforced by its application to the interpretation of XANES in these compounds as discussed in next section. The comparison of average Co-La/Pr and Co-Co bond lengths obtained from EXAFS and XRD given in table 2 indicates that the local structure is different from the average crystal structure.

## XANES

The experimental x-ray absorption spectra of LaCoO$_3$ and PrCoO$_3$ are shown in figure 7. The spectra have been normalized to unity at 100 eV above the edge. One can see nine features in the spectra denoted by A, B, C, D, E, F, G, H, and I. Two features A and B are pre-edge structures and earlier attributed to Co 1s $\rightarrow$ e$_g^\uparrow$ and e$_g^\downarrow$ transitions [17]. Some part of Co K-edge XANES in LaCoO$_3$ was also discussed in this work. However, the energy range in the present work is higher and the results are interpreted using DOS calculations under LDA. Rest of the features corresponds to transition of electron from Co 1s $\rightarrow$ 4p orbital hybridized with different orbitals. The pre-edge peaks have been plotted in the inset in zoomed scale. It is clear from the inset that the intensity of peak B slightly increases for PrCoO$_3$. The intensities of peaks D and G decrease whereas the intensity of peak I remains almost the same. Moreover, hump E and dip H are more prominent in PrCoO$_3$. The slight shift in the positions of features F, G, H, and I to the higher energy side is also seen.

The sensitivity of XANES to local structure has been shown by many workers [14-16]. To see the effect of lattice distortion on different XANES features we have calculated the absorption spectra for cubic and orthorhombic distortion. These spectra are plotted in upper panel of figure 8 along with the experimental data. Experimental spectra are rigidly shifted to match the position of main peaks with the calculated ones. It is clear from the figure that all the features except the two pre-edge features observed in the experimental spectra are also present in the calculated ones with slight deviation in the energy positions. The partial density of states (PDOS) of La 6p, Co 4p, and O 2p are also plotted in the lower panel of figure 8. By looking at the PDOS one can see that in the energy region between structures C and D the La 6p DOS is dominating; thus this region can be attributed to transition to Co 4p state strongly hybridized with La 6p state. However, in the energy region between features D and E the contribution of La 6p and O 2p DOS are almost the same. Therefore, this region is attributed to Co 4p state hybridized with La 6p and O 2p states. Similarly, region between structures E and F can be attributed to Co 4p hybridized with La 6p and O 2p states. In this region the contribution of La 6p is more in the hybridization in comparison to O 2p as the La 6p DOS is about 3 times more than that of O 2p. Structure G is attributed to the hybridized La 6p, Co 4p, and O 2p states, whereas structure I is attributed to Co 4p state hybridized with O 2p state.

It is evident from the upper panel of the figure 8 that the calculated spectra by taking Cubic and orthorhombic structures into account are representatives of the experimental spectra of LaCoO$_3$ and PrCoO$_3$, respectively. It may be noted that although the different features in the experimental spectrum match with the corresponding features in the calculated spectrum, there is a considerable difference in the post-edge intensities. This may be due to the fact that calculation of the matrix element was done by considering atomic states only rather than the solid-state wavefunction involving hybridization of Co 4p state with other states. Moreover, the absolute intensity of absorption spectrum depends on the effects like inelastic losses, extrinsic losses etc. It may be mentioned that such disagreement in experiment and calculation is quite known in literature and Rehr and Albers [30] have commented on it as one of the challenging problems in the field. The intensity of the main-edge peak D decreases for the orthorhombic structure in conformity with the observed decrease in the intensity of peak D for PrCoO$_3$. One can see overall small decrease in intensity in the region of peak G whereas in the region of peak I intensity increases. Moreover, orthorhombic distortion also shifts the positions of peaks G and I and dips F and H as observed experimentally for PrCoO$_3$. It is also clear from the calculated spectra that the features E and H are more prominent in the orthorhombic structure as seen experimentally. Thus calculated spectra by considering cubic and orthorhombic structures follow the experimental data of LaCoO$_3$ and PrCoO$_3$, respectively. It is thus seen that the orthorhombic distortion decreases the intensity of the main peak D and affects the intensity as well as position of the post-edge peaks. Our calculated spectra closely follow the changed behaviour of all the peaks except the peak I when one goes from LaCoO$_3$ to PrCoO$_3$. In the case of peak I the intensity of this peak increases for orthorhombic structure, which is opposite to the observed behaviour. This may be due to limitation of our calculations.

Finally, we comment on the increased intensity of the pre-edge peaks for PrCoO$_3$. As mentioned earlier the two pre-edge structures are attributed to Co 1s $\rightarrow$ e$_g^\uparrow$ and e$_g^\downarrow$ transitions. As seen from calculations, the e$_g$ orbitals are strongly hybridized with O 2p orbitals and their atomic character is thereby affected. Thus higher the admixture with O 2p, lower the Co e$_g$ atomic character and lower the intensity of the pre-edge peaks. This admixture is directly proportional to the average Co-O bond length. Thus the observed higher intensities of the pre-edge peaks in PrCoO$_3$ as compared to LaCoO$_3$ are in conformity with the EXAFS Co-O distances in these compounds rather than the XRD Co-O bond lengths. This is also as per known sensitivity of XANES to the local order.

**Summary and concluding remarks**

Room temperature Co K-edge EXAFS and XANES studies have been carried out on LaCoO$_3$ and PrCoO$_3$. The powder x-ray diffraction studies have shown the crystal structures of LaCoO$_3$ and PrCoO$_3$ to be rhombohedral and orthorhombic, respectively. In the rhombohedral structure all the six Co-O bond lengths should be equal, however, EXAFS analysis of LaCoO$_3$ data has revealed two different bond lengths making the CoO$_6$ octahedron distorted. The EXAFS studies of PrCoO$_3$ have also shown a similarly distorted $CoO_6$ octahedron. This distortion in octahedron at room temperature is attributed to the effect of $Co^{3+}$ ion in intermediate spin making it Jahn-Teller active. The higher shell fitting reveals that the Co-Co bond length in $LaCoO_3$ is more in comparison to that in $PrCoO_3$; however, higher values of Debye-Waller (DW) factor of this bond in $PrCoO_3$ indicates that this bond has more structural disorder in $PrCoO_3$ than in $LaCoO_3$. Such behaviour is also seen for Co-La/Pr bonds. This behaviour of bond lengths and DW factors indicates that structural disorder is playing an important role in deciding the increased resistivity of $PrCoO_3$ as compared to $LaCoO_3$. XANES studies of these compounds revealed the influence of crystal structure in deciding the intensities and energy positions of different features. The comparison of experimental and calculated spectra show that the intensity of main-edge and post-edge peaks decreases in orthorhombic structures as observed for $PrCoO_3$. The pre-edge peaks observed in both compounds are attributed to Co 1s $\rightarrow$ $e_g^{\uparrow}$ and $e_g^{\downarrow}$ transitions. The intensity of pre-edge peaks is seen to be more in $PrCoO_3$. This increase in intensity has been explained by using average Co-O bond length obtained from EXAFS rather than that from XRD. According to EXAFS the Co-O bond length is more for $PrCoO_3$ making Co $e_g$ orbitals less hybridized with O 2p orbitals. This decrease in overlap of these two orbitals enhances the atomic character of $e_g$ orbital resulting in increased intensity due to increase in matrix element. This result indicates the sensitivity of pre-edge features to the local distribution of surrounding atoms.

**Acknowledgements**

We would like to acknowledge R. Rawat for resistivity measurements. SKP thanks UGC-DAE CSR for financial support.

**Figure captions**

Figure 1    $k^2$- weighted EXAFS spectra $\chi(k)$ of $LaCoO_3$ (upper) and $PrCoO_3$ (lower).

Figure 2    Fourier Transform of $k^2\chi(k)$ for $LaCoO_3$ (upper panel) and $PrCoO_3$ (lower panel).

Figure 3    Fitted patterns of $LaCoO_3$ using *6 model* having six equal Co-O bonds and *2+4 model* having two short and four long Co-O bonds.

Figure 4    Fitted patterns of $PrCoO_3$ using *6 model* having six equal Co-O bonds and *2+4 model* having two short and four long Co-O bonds.

Figure 5    Higher shell fitted pattern of $LaCoO_3$ including Co-O, Co-La and Co-Co shells.

Figure 6    Higher shell fitted pattern of $PrCoO_3$ including Co-O, Co-Pr and Co-Co shells.

Figure 7    Normalized Co K-edge XANES spectra of $LaCoO_3$ and $PrCoO_3$ indicated by open circles and solid line, respectively. Inset shows zoomed two pre-edge structures denoted by A and B.

Figure 8    Normalized Co K-edge XANES spectra of $LaCoO_3$ and $PrCoO_3$ indicated by open circles and triangles, respectively and calculated spectra by taking cubic and orthorhombic structures denoted by solid and dotted lines, respectively are shown in the upper panel. The post-edge features in the calculated spectra corresponding to experimental ones are indicated by arrows. In the lower panel La 6p (solid lines), Co 4p (dashed lines), and O 2p (dotted lines) density of states per formula unit for different spin polarization are shown.

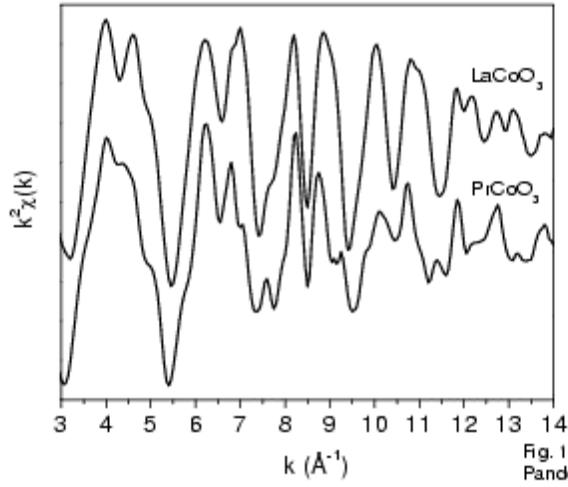

Fig. 1
Pandey et al.

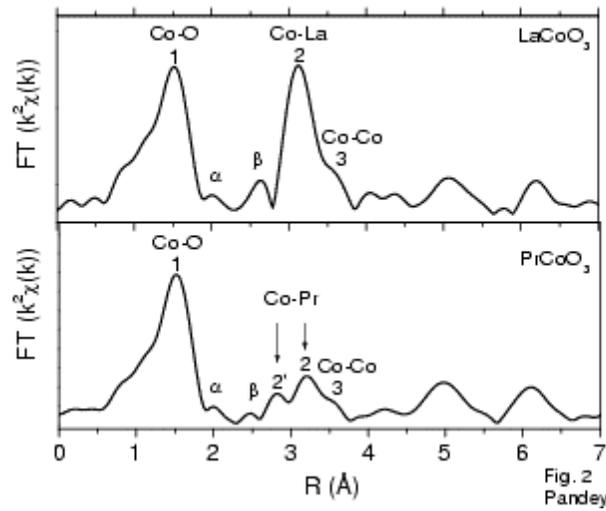

Fig. 2
Pandey et al.

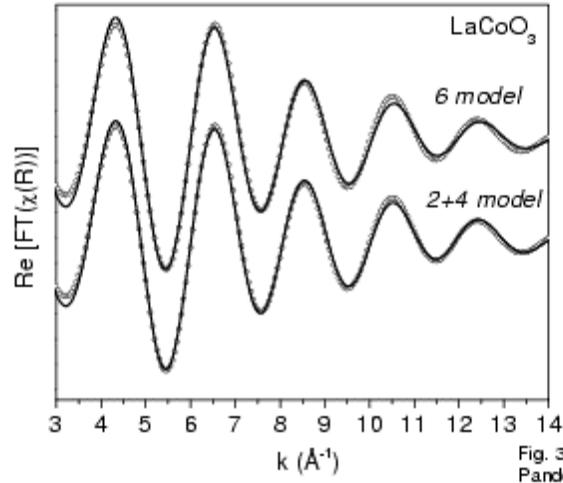
Fig. 3
Pandey et al.

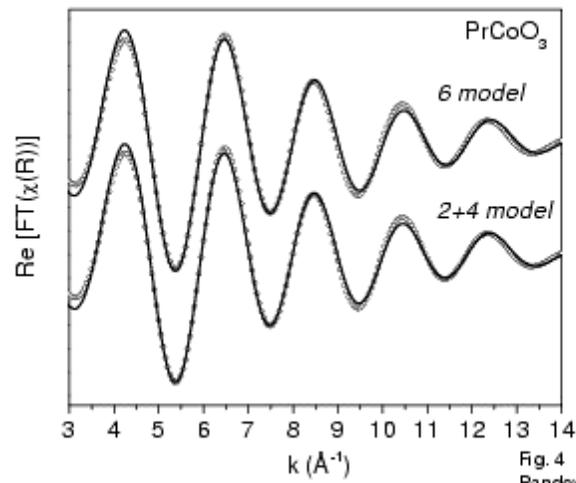
Fig. 4
Pandey et al.

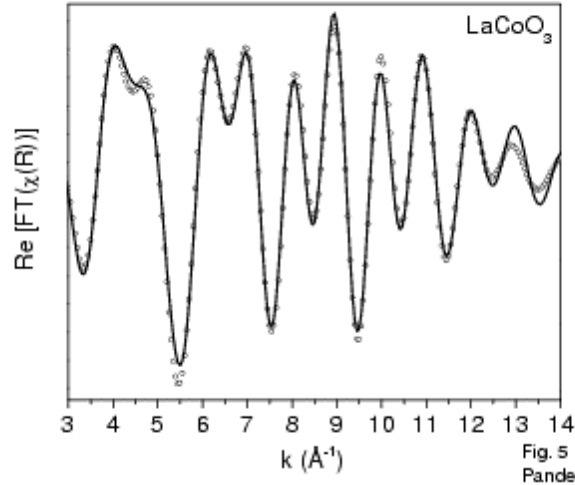

Fig. 5
Pandey et al.

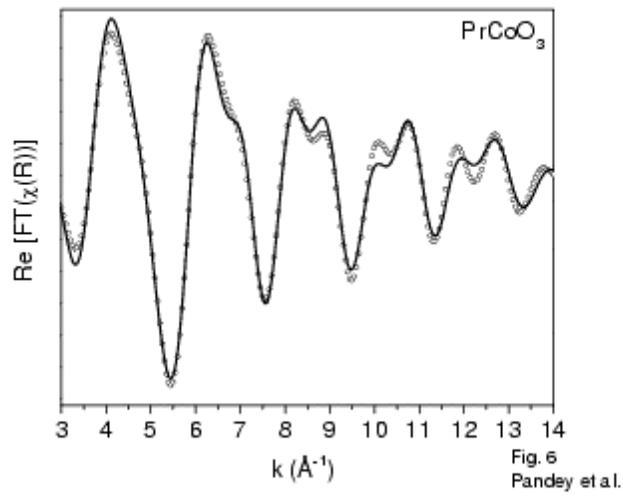

Fig. 6
Pandey et al.

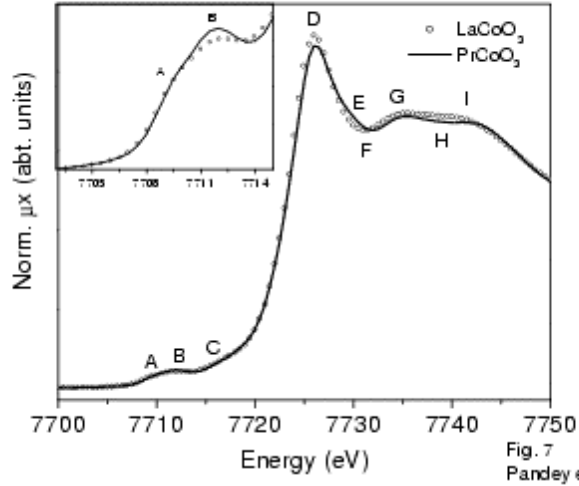

Fig. 7 Pandey et al.

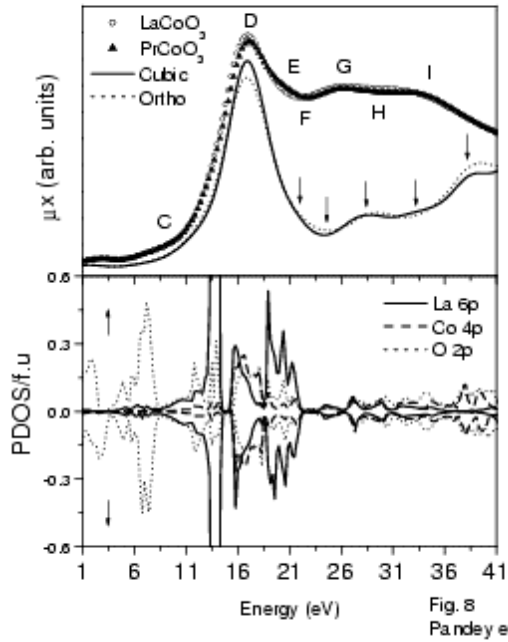

Fig. 8 Pandey et al.

Table 1

The Co-O bond lengths, Debye-Waller (DW) factors, and R-factors of LaCoO$_3$ and PrCoO$_3$ obtained from EXAFS analysis. The average Co-O bond lengths obtain from XRD is also given.

|  | LaCoO$_3$ | | | PrCoO$_3$ | | |
|---|---|---|---|---|---|---|
|  | Bond length (Å) | DW factor ($\times 10^{-3}$ Å$^2$) | R-factor | Bond length (Å) | DW factor ($\times 10^{-3}$ Å$^2$) | R-factor |
| *6 model* | 6× 1.914(1) | 4.1(2) | 0.0040 | 6× 1.921(2) | 3.5(4) | 0.0056 |
| *2+4 model* | 2× 1.863(6) 4× 1.928(2) | 4.5(8) 2.3(4) | 0.0025 | 2× 1.886(8) 4× 1.942(4) | 3.0(1.2) 2.7(6) | 0.0048 |
| Average | 6× 1.906 | | | 6× 1.923 | | |
| XRD | 6× 1.936(5) | | | 6× 1.926(4) | | |

Table 2

The Co-La/Pr and Co-Co bond lengths, Debye-Waller (DW) factors, and R-factors of LaCoO$_3$ and PrCoO$_3$ obtained from EXAFS analysis. The average Co-La/Pr and Co-Co bond lengths obtain from XRD are also given.

| Compound | Type of bonds | EXAFS | | | XRD |
|---|---|---|---|---|---|
|  |  | Bond length (Å) | DW factor ($\times 10^{-3}$ Å$^2$) | R-factor | Bond length (Å) |
| LaCoO$_3$ | Co-La Co-Co | 2× 3.251(17) 6× 3.334(5) 6× 3.894(9) | 5.0(1.5) 4.6(5) 6.5(7) | 0.012 | 2× 3.273(1) 6× 3.325(1) 6× 3.825(1) |
| PrCoO$_3$ | Co-Pr Co-Co | 2× 3.128(16) 6× 3.308(7) 6× 3.818(8) | 7.0(2.6) 10.0(1.1) 9.0(1.0) | 0.015 | 2× 3.140(8) 6× 3.331(3) 6× 3.788(1) |